\documentclass{elsart}
\usepackage{graphicx}
\begin{document}
\begin{frontmatter}
\title{Critical exponents and phase transition in gold nuclei fragmentation at
energies 10.6 and 4.0 GeV/nucleon}
\author{D.Kudzia},
\author{B.Wilczy\'nska},
\author{H.Wilczy\'nski}
\address{Institute of Nuclear Physics, Radzikowskiego 152, 31-342 Krak\'ow, 
Poland}

\begin{abstract}
An attempt to extract critical exponents $\gamma$, $\beta$ and
$\tau$ from data on gold nuclei fragmentation due to interactions
with nuclear emulsion at energies 4.0 A GeV and 10.6 A
GeV is presented. Based on analysis of Campi's 2nd charge moments, two subsets 
of data at each energy are selected from the inclusive data, corresponding to 
'liquid' and 'gas' phases. The extracted values
of critical exponents for the selected data sets are in
agreement with predictions of 'liquid-gas' model of phase
transition.
\end{abstract}

\end{frontmatter}

\section{Introduction}

Multifragmentation, a breakup of an excited nucleus into many
intermediate mass fragments, has been discussed for almost twenty years
 in terms of statistical mechanics; possible critical
behavior was investigated. In its ground state nuclear matter behaves like
a liquid. Mean field theory simulations \cite{Sil01} predict that
nuclear equation of state resembles that of Van der Waals gas.
Therefore existence of phase transition, spinodal instabilities
and critical point are expected. However, nature of the possible
phase transition is still under debate. Existing experimental data
do not lead to a conclusive answer 
\cite{Pochodzalla95,Hauger96,Nebauer01,Kleine02}.

In many papers it was shown (e.g.\cite{Hauger96})
that nuclear interactions undergo two stages. During the first
stage, prompt nucleons are emitted from the colliding system and they
carry out a large amount of available kinetic energy. They result
from quasi-elastic and nonelastic collisions of projectile and
target nucleons. Immediately after the collision, the remnant of
the nucleus is in an excited state with temperature $T_i$. At the
second stage, the excited remnant expands and cools evolving into
neighborhood of the  critical point on the temperature--density
plane \cite{Hauger98}. Then the hot and decompressed nucleon gas
with temperature $T_f$ condenses into many fragments. This last
process is the multifragmentation. It was also shown that total
charged fragment multiplicity is proportional to the temperature
of the colliding system, both to $T_i$ and $T_f$ \cite{Hauger98}.

In the 90-ties there were many attempts to extract critical
exponents in nuclear fragmentation from the experimental data, e.g. 
\cite{Gilkes94,Cherry95,Adamovitch98,Adamovitch99}. A method of
charged moments invented by Campi \cite{Campi86} and supported by
percolation theory \cite{Stauffer92} was commonly applied. These attempts did 
not
take into account the fact that it is only the remnant of the nucleus
that undergoes multifragmentation process, so all prompt
particles (participating in the nucleus-nucleus collision)
 were included in the analysis. This approach raised a
critique \cite{Bauer95,Gilkes95}, that pointed out that:
inclusion of prompt particles, assumption that fragment
multiplicity is directly proportional to the temperature of the
fragmenting system and using fragment charges
$Z_i$ instead of fragment masses $A_i$ are not justified and can
significantly change the resulting values of critical exponents.

Recently, the EOS collaboration \cite{Elliott011} calculated critical
exponents based on a high-statistics sample of fully reconstructed events  from
fragmentation of Au nuclei with energy 1.0 GeV/nucleon, taking
into account the above criteria and using fragment masses, not charges. The 
obtained
results are in agreement with the previously extracted values of the critical 
exponents
\cite{Gilkes94}. Therefore the exclusion of prompt particles and
performing analysis based on moments of mass distribution, instead of
charge distribution, does not impact resulting values. This can be
understood based on the fact that a large majority of prompt particles
are fragments with charge $Z=1$. Their impact on the second
charge moments (that take square of the charges) is much smaller than
that of heavier fragments.

In this paper we present the analysis of data coming from
interactions of projectile gold nuclei of primary energies 4.0 and
10.6 GeV/nucleon with nuclear emulsion target.

\section{The experiment}

Stacks of BR-2 emulsion pellicles were irradiated
with gold ion beam at the AGS accelerator at Brookhaven National
Laboratory. The stacks were oriented so that the beam was parallel
to the pellicles. Interactions were found during a microscope scanning
along the primary tracks in order to obtain a sample with minimum
detection bias. The two data sets consist  of 448 events
at 4.0 GeV/nucleon and 884 events at 10.6 GeV/nucleon.
In each event analyzed, multiplicities and emission angles of all
produced particles and fragments of
colliding nuclei were measured. In addition, charges of projectile
nucleus fragments were determined. Singly-charged particles
(released protons and produced pions) were distinguished unambiguously from
heavier
fragments. Charges of heavier fragments ($Z \ge 2$) of the projectile were
measured
using a photometric method with a CCD camera \cite{Kudzia98}.

The experimental method of identification of prompt particles
(i.e. nucleons directly participating in the collision) in
emulsion experiments is not available. We can however estimate the number
of prompt particles as a difference between the total number of
emitted protons $N_{prot}$ and the number of spectator protons
$N_{spec}$. $N_{prot}$ is determined using charge balance of the projectile
fragments.
 We have estimated $N_{spec}$
as the number of singly charged particles that are
emitted at the angle $\theta < 2\theta_0$, where $\theta_0$ is the
average proton emission angle given by relation $ \theta_0 = 0.12/p$,
and $p$ is the momentum of the nucleon prior to emission
\cite{Powell59}.

In this paper critical exponents were extracted using the Campi's charge 
moments method. The charge moments in our analysis were calculated in two 
different ways.
One of them is to find non-normalized charge moments taking
into account all heavy ($Z>2$) fragments, alpha particles and  protons
emitted from the vertex of interaction following Gilkes at al
\cite{Gilkes94}. The other way is for normalized charge moments and
excludes prompt protons following Elliott et al. \cite{Elliott011}. The two
ways were used to
detect possible inconsistencies coming from the choice of the method.

\section{Charge moments}

Following Campi \cite{Campi86,Campi88}, we define the total charged
fragment multiplicity $m$, as $m = N_f + N_{\alpha} + N_{prot}$,
where $N_f$ denotes the number of fragments with charge $Z \ge 3$,
$N_{\alpha}$ is the number of emitted alpha particles and $N_{prot}$
is the number of emitted protons.

The distance from the critical point for a given event
can be properly measured by a difference between multiplicity $m$
and the multiplicity at the critical point
$m_c$. The validity of this assumption rests on a linear dependence
between temperature of the system $T$ and the total multiplicity
$m$, that was shown to be valid by Hauger at al. \cite{Hauger96}.
Therefore we introduce variable $\epsilon$:
\begin{equation}
\label{eq:1}
    \epsilon = m - m_{c},
\end{equation}

Fragments of a given charge, $Z_f$, were counted on event-by-event
basis to determine fragment charge distributions $N_{Z_f}$.
The normalized charge distribution was defined as $n_{Z_f} =
N_{Z_f}/Z_{A_0}$, where $Z_{A_0}$ is charge of the nucleus
remnant undergoing fragmentation, given by a sum of charges of
all fragments $Z_f \ge
2$ plus total charge of spectator protons.

In order to compare our results with theoretical calculations
(the Fisher model), mass distributions should be used
and should be normalized to the remnant mass $A_0$ 
rather than to its charge $Z_{A_0}$. In emulsion experiments, fragment mass
is not measurable; therefore it was assumed that fragment charge
$Z_f$ is proportional to its mass $A_f$. It was also assumed that on
average, charge and mass distributions are equal $N(Z_f)= N(A_f)$.
In fact, the same assumption is made also by other authors,
at least in
specific  ranges of charges. This is dictated by the difficulty in
measuring masses of all heavier fragments emerging from the
interaction vertex. In case
of electronic experiments masses of fragments
 only in some small
mass range are measured. 
Therefore, we have to rely on the assumption of
proportionality between mass and charge.

Following Campi \cite{Campi86}, we define the k-th moment of charge
distribution as:
\begin{equation}
\label{eq:2}
    M_k(\epsilon) = \sum n_{Z_f}(\epsilon)Z_{f}^{k},
\end{equation}
where the sum extends over all charged fragments except prompt protons
in the 'gas phase', that is for $\epsilon > 0$. In the 'liquid
phase',where $\epsilon < 0$, in addition the fragment with highest
charge $Z_{max}$ is omitted from summation.
The above procedure is motivated by the Fisher model. The bulk liquid
of infinite volume is excluded from calculation on the 'liquid'
phase. Similar procedure is carried out in percolation models, where
percolating cluster of infinite size is excluded from calculation
on the 'liquid' phase.

With the above assumptions, and  based on the Fisher
model, in the thermodynamic limit the following
relations are valid \cite{Elliott011}:
\begin{equation}
\label{eq:3}
    M_2(\epsilon) \sim |\epsilon|^{-\gamma},
\end{equation}
\begin{equation}
\label{eq:4}
    n_{Z_f}(\epsilon) \sim Z_{f}^{-\tau} \;\; \mbox{for} \;\; \epsilon = 0
\end{equation}
In addition, Bauer et al \cite{Bauer88} have shown that:
\begin{equation}
\label{eq:5}
    Z_{max}(\epsilon) \sim \epsilon^{\beta} \;\; \mbox{for} \;\; \epsilon < 0,
\end{equation}
In these equations $\beta$, $\gamma$  and $\tau$ are the critical
exponents. $Z_{max}(\epsilon)$ is the average charge of the largest
fragment for a given value of the distance from critical
point $\epsilon$. The above relations are valid in the
neighborhood of the critical point. Far away from the critical
point, the behavior of the system is dominated by the mean field
regime and these relations are not followed. Very close to the
critical point, on the other hand, finite size effects come into
play and $M_2$ does not raise to infinity with $\epsilon$
approaching 0 (Equation \ref{eq:3}), but achieves some maximum value.
The critical exponents  $\beta$, $\gamma$ and  $\tau$ are not
independent.  A scaling relation between them exists:
\begin{equation}
\label{eq:6}
    \tau = 2+ \frac{\beta}{\beta+\gamma}
\end{equation}

Practical calculations of the moments of charge distributions were
based on averaging in the small bins of multiplicity $m$ of event-by-event
distributions \cite{Elliott011,Gilkes94}:
\begin{equation}
\label{eq:7}
    \langle M_k(\epsilon)\rangle = \frac{1}{N}\sum M_{k}^{i}(\epsilon)
                    =  \frac{1}{N} \sum_i
                   ( \sum_{Z_f} n_{Z_f}^{i}(\epsilon)Z_{f}^{k})
\end{equation}
where $i$ is the index of an event,  $N$ denotes the total number of
events in a given small range of $\epsilon$, $M_{k}^{i}$ is a
charge distribution moment for event $i$.

\section{Fluctuations close to the critical point}

One of the basic effects of the second order phase transition is
the appearance of significant fluctuations in the
neighborhood of the critical point, in a small range of
temperatures $T$ (or other parameter measuring distance from the
critical point). Fluctuations grow as the critical point is
approached and appear at increasingly large scales. In the case
of a normal liquid, the effect is known as the critical opalescence:
fluctuations of density
of the liquid and sizes of gas bubbles result from vanishing
latent heat of the phase transition as the critical point is
approached. In the Fisher model of nuclear fragmentation, this is
reflected by the divergence of the isothermal compressibility
$\kappa_{T}$ at the critical temperature $T_{c}$: small variations
of the pressure result in big density changes.
In the neighborhood of the critical point, the volume and surface
terms of the Gibbs free energy of fragment formation vanish and
the fragment distribution is dominated by the power law of Equation \ref{eq:4}
\cite{Fisher67}.

In case of multifragmentation, fluctuations can be analyzed using
moments of charge distributions. Campi suggested 
a variable $\gamma_2$ linearly dependent on variance $\sigma^2$ of
the charge $\langle Z \rangle$ distribution:
\begin{equation}
\label{eq:3_28}
    \gamma_2 =  \frac{M_2 M_0}{M_{1}^{2}} = 1 +
    \frac{\sigma^2}{\langle Z \rangle ^2},
\end{equation}

\begin{figure}
\includegraphics[scale=0.48,bb= 13 280  850 810]{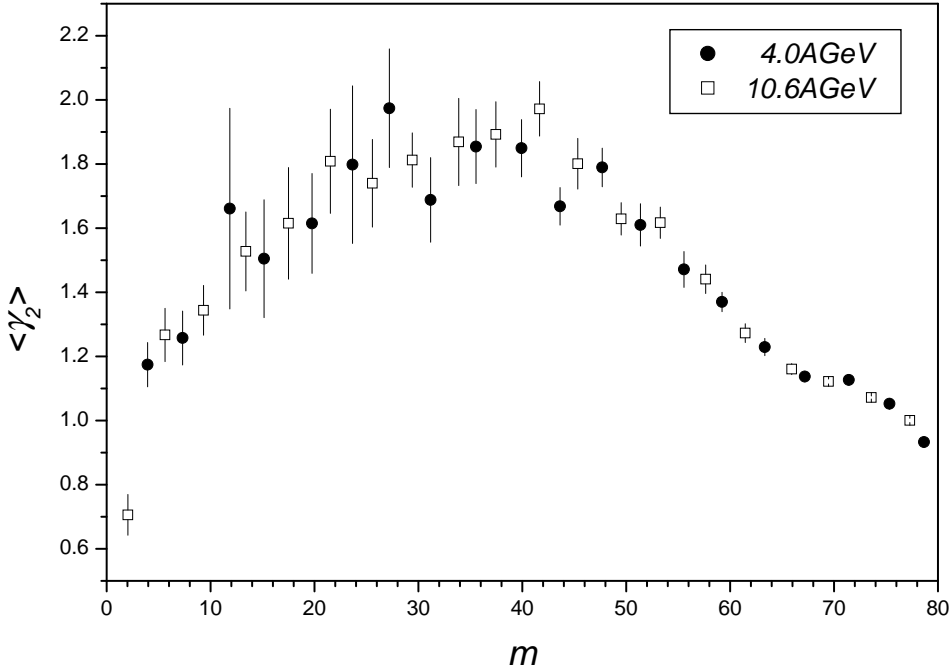}
\caption{ $\langle \gamma_2\rangle$ as a function of total charged
fragment multiplicity $m$ at energies 4.0 and  10.6 GeV/nucleon. The error bars
show the dispersion of $\gamma_2$ in each bin.}
\label{Fig_1}
\end{figure}

Figure \ref{Fig_1} shows $\langle\gamma_2\rangle$ as a function of
multiplicity $m$ for energies 4.0 and 10.6 GeV/nucleon. As
critical value of multiplicity $m_c$ is not identified at
this stage, in all above calculations charge moments were calculated
with $Z_{max}$ excluded.
For multiplicities smaller than $m \simeq 30$,
$\langle\gamma_2\rangle$ grows monotonically with multiplicity,
for multiplicities larger than $m \simeq 40$, it
falls down monotonically.
Fluctuations of $\gamma_2$, reflected in large dispersion, are largest for
multiplicities between $10$ and $40$.
It is commonly assumed that strong maximum of $\langle
\gamma_2\rangle$ results from undergoing a phase transition at
critical point multiplicity $m = m_c$. The region of
multiplicities $ m < m_c$ is called a 'liquid' phase, and events
with $ m > m_c$ are called to be in 'gas' phase. However, the presence of
maximum of $\langle \gamma_2\rangle$ is not a conclusive argument
for appearance of phase transition. It was shown
\cite{Elliott011} that such maximum is observed also in systems
that do not undergo a phase transition. Therefore Figure
\ref{Fig_1} will be treated as only a hint for possibility of phase
transition in a specific range of multiplicities. It is also worth
mentioning that $\gamma_2$ fluctuations at energies 4.0 and 10.6A GeV
agree well with each other.

\begin{figure}
\includegraphics[scale=0.48,bb= 13 280  850 810]{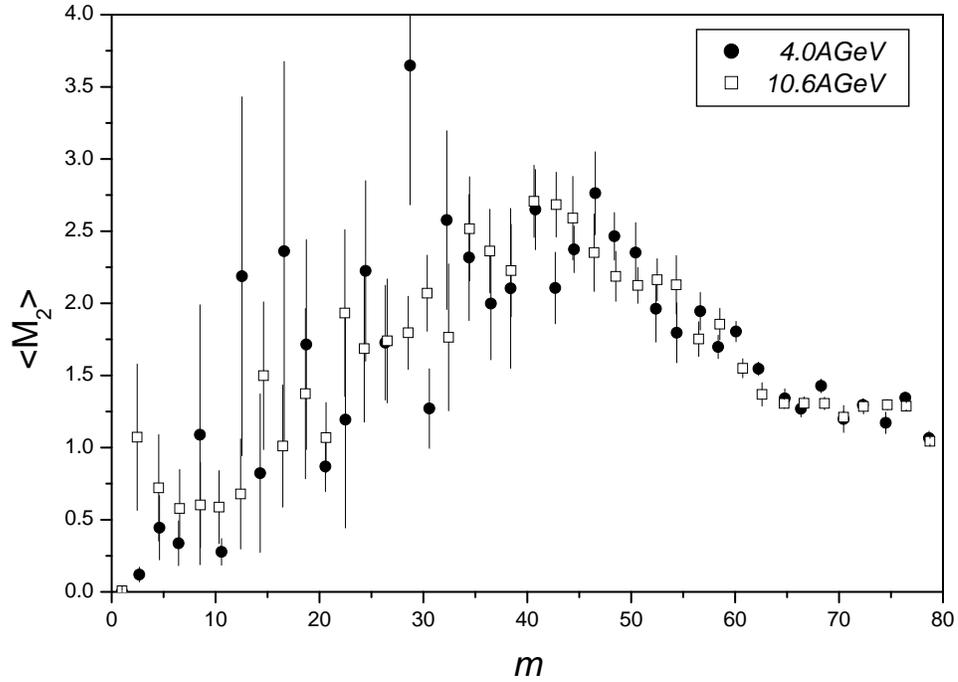}
\caption{Mean value of the second charge moment $\langle M_{2}\rangle$
as a function of multiplicity  $m$ for energies 4.0 and 10.6 GeV/nucleon. The 
error bars
show dispersion of $M_2$ values for each bin.}
\label{Fig_2}
\end{figure}

\begin{figure}
\includegraphics[scale=0.48,bb= 13 280  850 810]{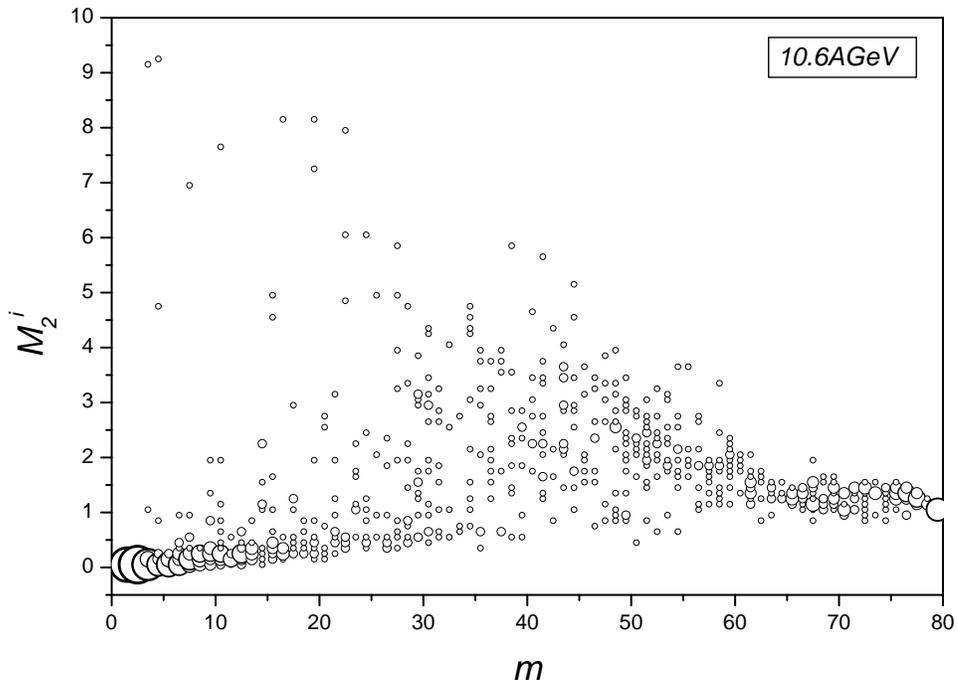}
\caption{Scatter plot of second charge moments for individual
events $M_{2}^{i}$ as function  multiplicity $m$ for energy 10.6
GeV/nucleon.}
\label{Fig_3}
\end{figure}

In order to closer analyze the fluctuations, the mean values of second
moment of charge distribution $\langle M_{2}\rangle$ were plotted
versus multiplicity in Figure \ref{Fig_2}. Similar characteristics
to that of $\langle\gamma_2\rangle$ is seen. Large dispersion of $M_2$,
especially in the 'liquid' phase reflects strong fluctuations of
experimental values.

\begin{figure}
\includegraphics[scale=0.48,bb= 13 280  850 810]{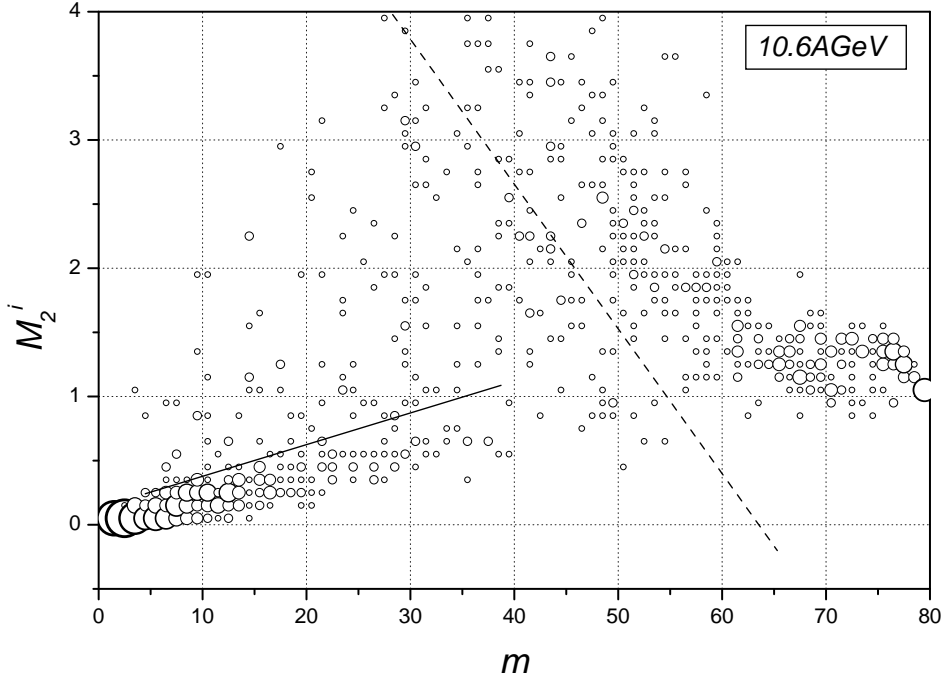}
\caption{Selection of 'liquid' and 'gas' groups of events in the $M_2^i$---$m$ 
scatter plot.}
 \label{Fig_4}
\end{figure}

\begin{figure}
\includegraphics[scale=0.48,bb= 13 280 850 810]{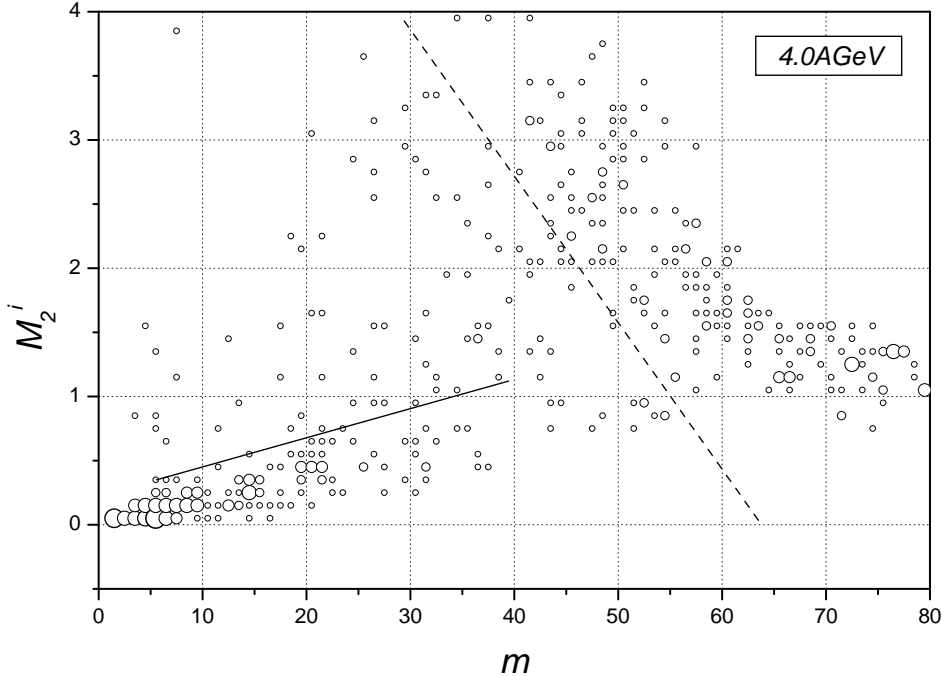}
\caption{Second charge moment $M_{2}^{i}$ for individual events
vs. multiplicity $m$ for energy 4.0 GeV/nucleon. Events selection
criteria .}
 \label{Fig_4a}
\end{figure}

Figure \ref{Fig_3} shows a scatter plot of second charge moments
for individual events $M_{2}^{i}$ versus multiplicity
$m$ at energy 10.6 GeV/nucleon. Strong fluctuations of
$M_{2}^{i}$ for multiplicities $15 \le m \le 45$ are clearly
visible, which result in large dispersion of $M_2$ values (shown as error bars
in Figure \ref{Fig_2}). For systems undergoing a second order phase transition,
it is
expected that fluctuations rise in the neighborhood of the
critical point. Therefore observation of large fluctuations quite
far away from expected critical region should be attributed to a
different physical phenomenon. For example, events with values of
$M_{2}^{i} > 4$ and multiplicity  $m < 20$ are fission-like (only 2 heavy
fragments plus alphas and protons).
Two distinct groups of experimental
points are seen in Figure \ref{Fig_3}. One of these groups consists of
events with
multiplicity smaller than $m=35$ and $M_2 < 0.9$, the other --- of events
with multiplicities larger than $m=30$ and largest $M_2$ values for a given
$m$. The former group, with small multiplicities and small $M_2^i$ suggests
that the fragment charge distribution must be dominated by one heavy fragment
-- i.e. the expected characteristics of the 'liquid' phase. On the other hand,
the latter group of events, with large multiplicity and larger $M_2$, suggests
existence of a larger number of small fragments, thus it resembles the
expected characteristics of the 'gas' phase. The events which do not belong to
either of these groups may be thought of as nuclei in a 'mixed' phase (see 
Section 7).
The 'liquid' and 'gas' groups were selected from experimental data
for further analysis. Figure \ref{Fig_4} shows the selection
criteria. The 'liquid' group of events consist of those with multiplicity
smaller than $m=35$ located below the solid line, while the 'gas' group
consists of events with $m > 30$ and located above and to the
right of the dashed line. Similar selection was performed for data
with energy 4.0 GeV/nucleon -- see Figure \ref{Fig_4a}.

\vspace{5mm}

\begin{figure}
\includegraphics[scale=0.48,bb= 13 280  850 810]{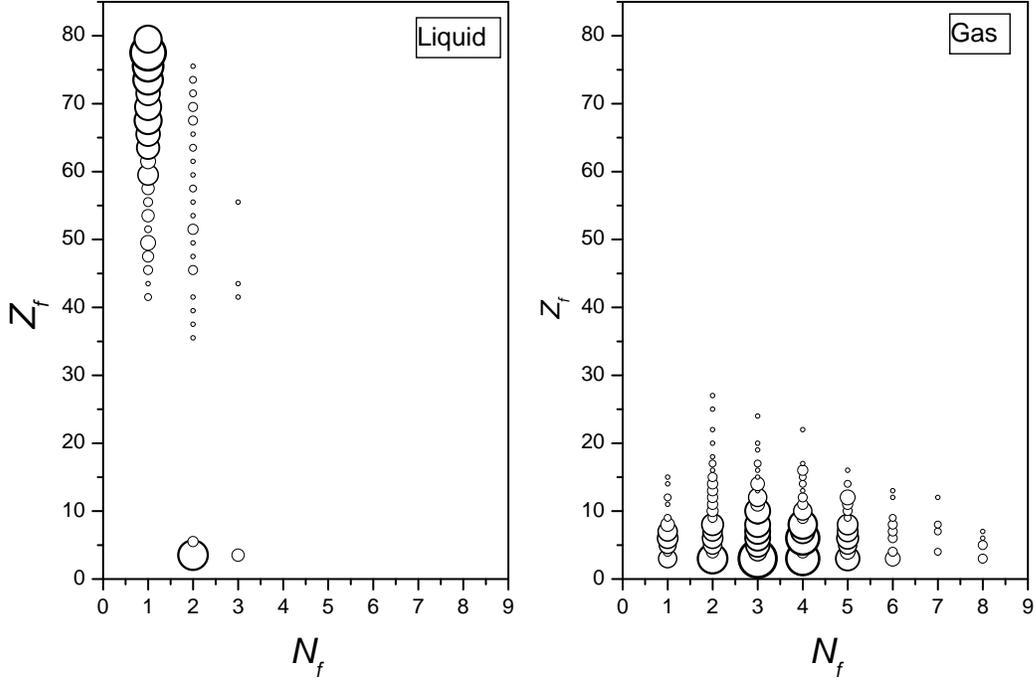}
\caption{Charge distributions in events with different heavy fragment 
multiplicity $N_f$
 for the two selected groups of experimental
data at energy 10.6 GeV/nucleon. }
 \label{Fig_5}
\end{figure}

Let us look at these two selected groups in order to check
if they really can be treated as 'liquid' and 'gas' phases. For example,
Figure \ref{Fig_5} presents charge distributions as a function of heavy 
($Z>2$) fragment multiplicity, $N_f$, for the two selected
groups of data. Fragment charge distributions
are clearly different in both groups.  The charge distribution for the
group of low multiplicity events consists mainly of events
with one heavy fragment emitted ($Z \ge 45 $). If the second
fragment is emitted, it has a small charge ($Z \le 3$). Therefore, we
will refer to
this group of data as 'liquid' as for these events one
large fragment remains after collision in analogy to the large
drop of liquid. In most cases there are only alphas and protons accompanying
the large heavy fragment. The second group of events has totally
different charge distribution that consists of events with number
of emitted heavy fragments ranging from 1 to 9 and charges of fragments
$Z \le 25$. The majority of fragments in this group has
charges $Z \le 15 $. We will call this group of
events a 'gas' as it consists of events with many light fragments
emitted. The analysis presented in subsequent sections presents
results on these selected sets of data unless
explicitly stated otherwise.

\section{Critical exponent $\gamma$ }

In order to extract the critical exponent $\gamma$ and the critical
value of multiplicity $m_c$ the method known as 'gama matching'
was used \cite{Bauer88,Elliott011}. The outline of this method is the
following.
A trial value of the critical multiplicity $m_c$ is chosen. For
a given $m_c$, a distribution of mean values of the moment
$\langle M_2(\epsilon) \rangle$  is determined as a function of
distance from the critical point $\epsilon = m-m_c$. Then the
ranges in $\epsilon$ are chosen to fit the power law (\ref{eq:3}) to the
experimental data, separately for the 'gas'
and 'liquid' phases. With fitting
boundaries determined, the linear fit to the $ln\langle M_2(\epsilon) \rangle$
versus $ln|\epsilon|$ is made to extract values of the slope $\gamma$
separately for 'gas' and 'liquid' phases.

\begin{figure}
\includegraphics[scale=0.48,bb= 13 280  850 810]{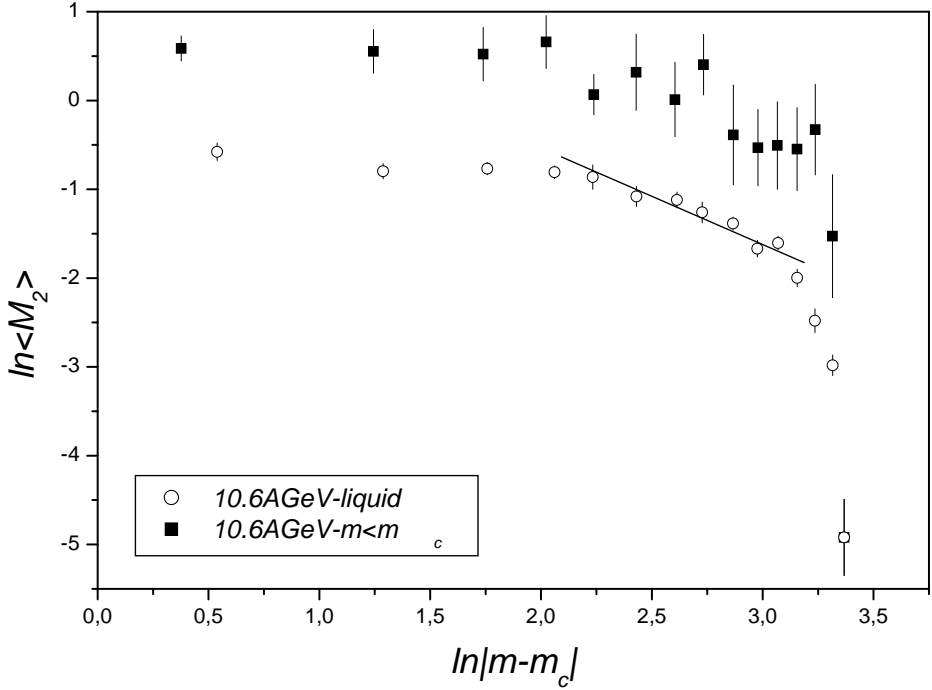}
\caption{Mean value of second charge moment $\langle M_{2}\rangle$
for a 'liquid' phase as a function of multiplicity $m$ for energy
10.6 GeV/nucleon. Full squares show data for all events with $m<m_c$, open
squares represent data for selected 'liquid' events only. The line shows
the fit used to determine critical exponent $\gamma$.}
\label{Fig_9}
\end{figure}

\begin{figure}
\includegraphics[scale=0.48,bb= 13 280  850 810]{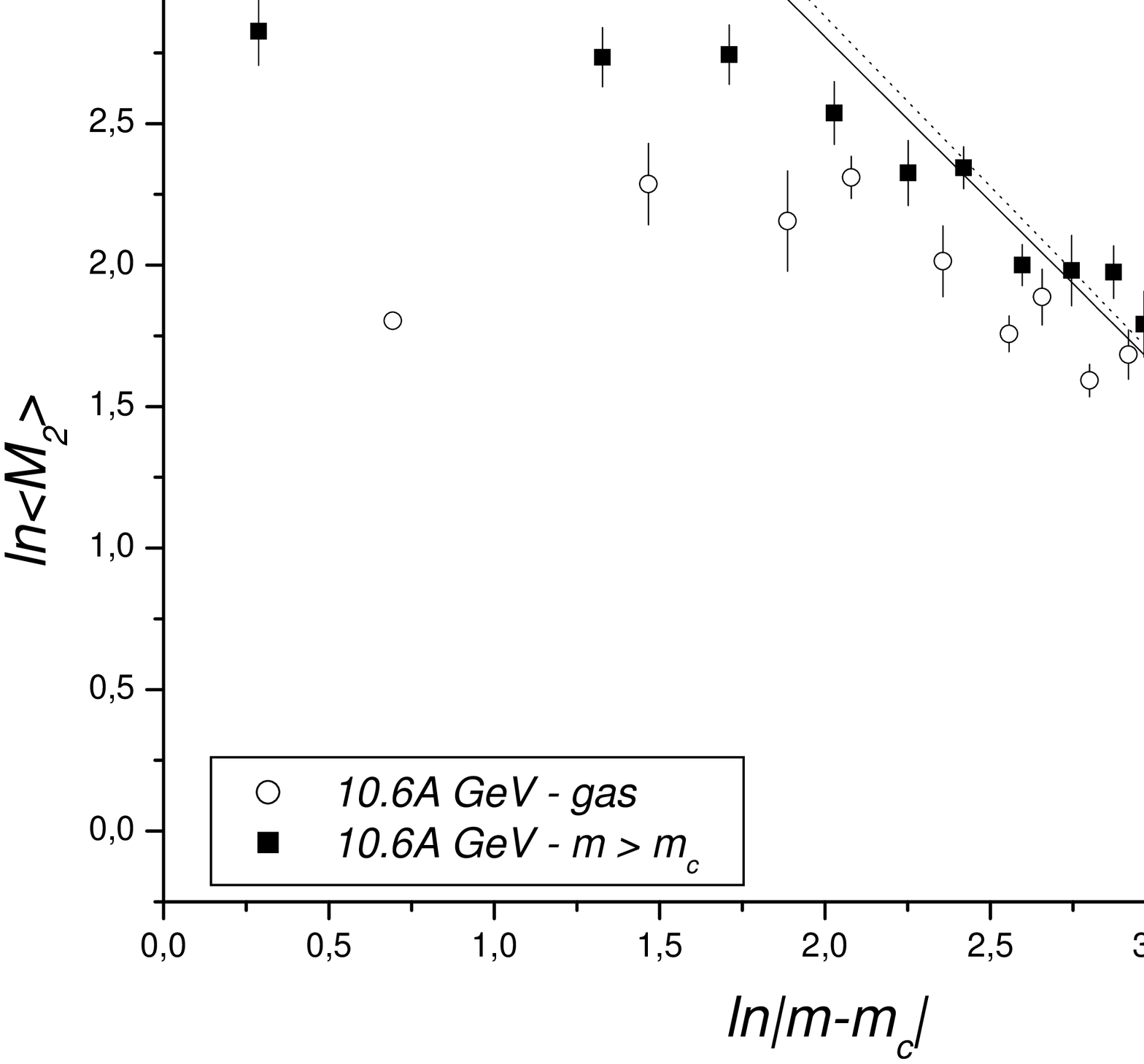}
\caption{Mean value of second charge moment $\langle M_{2}\rangle$
for a 'gas' phase as a function of multiplicity $m$ for energy
10.6 GeV/nucleon. Full squares show data for all events with $m>m_c$, open
squares represent data for selected 'gas' events only. Plotted lines
show fits to the 'gas' data (solid line) and $m>m_c$ data (dotted line).}
\label{Fig_10}
\end{figure}

The fitting boundaries for the
'liquid' phase  are easily determined.
Figure \ref{Fig_9} presents a comparison of second moments of charge
distribution for the selected 'liquid' phase and all available
experimental data at $m<m_c$.  It is seen that the
selection results in appearance of a very clear region of power law
dependence.

\begin{figure}
\includegraphics[scale=0.48,bb= 13 280  850 810]{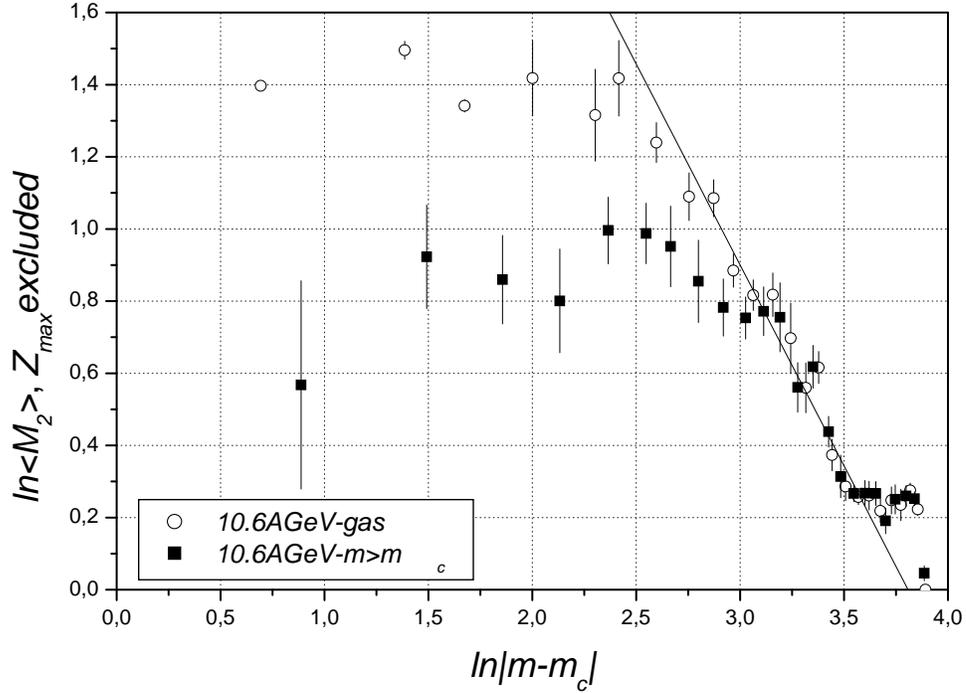}
\caption{Mean value of the second charge moment $\langle M_{2}\rangle$
for a 'gas' phase with exclusion of the largest charge $Z_{max}$
as a function of multiplicity $m$ for energy 10.6 GeV/nucleon.}
\label{Fig_8}
\end{figure}

Similar comparison for the 'gas' phase is
shown in Figure \ref{Fig_10}. The
impact of selection of the 'gas' phase is not significant in this case.
The region of the power law dependence  is not clearly
seen. This may be due to the inclusion of the charge of the largest
fragment $Z_{max}$ in the 'gas' phase. The largest charge dominates
the second moment, shifts it to the higher values and makes the power
law less visible, probably due to the finite size effect.
Thus, it may be interesting to check if the exclusion of $Z_{max}$ also in the
'gas' phase reveals a clear range of power law dependence of $<M_2>$ versus 
$\epsilon$.
 The $M_2$
 moment  for the 'gas' phase computed
without $Z_{max}$ is shown in Figure \ref{Fig_8}. Open circles
represent moments computed for 'gas' group of
data, filled circles represent those for all experimental
data at $m>m_c$, without selection. In both cases the region of power law
behavior is now clearly visible. This region is much larger for selected 'gas'
data compared to $m>m_c$ data, so it is now clear in which region of
$\epsilon$ the fit should be performed. In other words, Figure \ref{Fig_8} was
used as a guideline in determining
fitting region in the 'gas' phase in the 'gamma matching procedure'.
It is interesting that the value of $\gamma_{gas} = 1.14 \pm 0.05$
determined from moments of charge distribution with $Z_{max}$
excluded is very close to the value determined by the 'gamma matching
procedure' (i.e. with $Z_{max}$ included).

 The above procedure was repeated for several
trial values of the
critical point $m_c$. The value of the critical exponent $\gamma$
and critical point $m_c$ is found by demanding that
$|\gamma_{gas} - \gamma_{liquid}|$ takes a minimum value and
$\gamma_{gas}$ and $\gamma_{liquid}$ agree with each other within
statistical errors. Trial values of the critical point were
selected in the range $25 \le m_c \le 35$.

\begin{table}
\caption{Critical exponent $\gamma$ for 'gas' and 'liquid' phases for
different choices of trial value of critical point $m_c$ for
energy 10.6 GeV/nucleon used in gamma matching procedure. Normalized
moments}
\label{Tab_gamma10Gev-norm}
\begin{center}
\begin{tabular}{|c|c|c|c|}
\hline
trial $m_c$ & $\gamma_{liquid}$ & $\gamma_{gas}$& 
$|\gamma_{liquid}-\gamma_{gas} |$ \\
\hline
$26$ & $0.80  \pm 0.00$ & $1.45  \pm 0.06$ & $0.63  \pm$ 0.10\\
$28$ & $0.96  \pm 0.11$ & $1.32  \pm 0.05$ & $0.35  \pm 0.12$\\
$29$ & $1.04  \pm 0.11$ & $1.26  \pm 0.05$ & $0.22  \pm 0.12$\\
$30$ & $1.12  \pm 0.12$ & $1.20  \pm 0.05$ & $0.08  \pm 0.13$\\
$31$ & $1.20  \pm 0.13$ & $1.14  \pm 0.05$ & $0.05  \pm 0.14$\\
$32$ & $1.27  \pm 0.14$ & $1.08  \pm 0.05$ & $0.19  \pm 0.15$\\
$34$ & $1.42  \pm 0.15$ & $0.96  \pm 0.04$ & $0.46  \pm 0.16$\\
\hline
\end{tabular}
\end{center}
\end{table}

\begin{table}
\caption{Critical exponent $\gamma$ for 'gas' and 'liquid' phases for
different choices of trial value of critical point $m_c$ for
energy 10.6 GeV/nucleon used in gamma matching procedure. Non-normalized
moments.}
\label{Tab_gamma10Gev}
\begin{center}
\begin{tabular}{|c|c|c|c|}
\hline
trial $m_c$ & $\gamma_{liquid}$ & $\gamma_{gas}$& $|\gamma_{gas} -
\gamma_{liquid}|$ \\
\hline
$26$ & $0.78  \pm 0.07$ & $1.30  \pm 0.11$ & $0.48  \pm$ 0.13\\
$28$ & $0.93  \pm 0.09$ & $1.22  \pm 0.10$ & $0.20  \pm 0.14$\\
$29$ & $1.02  \pm 0.10$ & $1.17  \pm 0.10$ & $0.15  \pm 0.14$\\
$30$ & $1.09  \pm 0.10$ & $1.13  \pm 0.10$ & $0.03  \pm 0.14$\\
$31$ & $1.17  \pm 0.10$ & $1.08  \pm 0.10$ & $0.09  \pm 0.14$\\
$32$ & $1.24  \pm 0.12$ & $1.03  \pm 0.09$ & $0.21  \pm 0.14$\\
$34$ & $1.39  \pm 0.13$ & $0.94  \pm 0.08$ & $0.45  \pm 0.15$\\
\hline
\end{tabular}
\end{center}
\end{table}

\begin{table}
\caption{Critical exponent $\gamma$ for 'gas' and 'liquid' phases for
different choices of trial value of critical point $m_c$ for
energy 4.0 GeV/nucleon used in gamma matching procedure. Normalized
moments.} \label{Tab_gamma4Gev-norm}
\begin{center}
\begin{tabular}{|c|c|c|c|}
\hline
trial $m_c$ & $\gamma_{liquid}$ & $\gamma_{gas}$& $|\gamma_{gas} -
\gamma_{liquid}|$ \\
\hline
$26$ & $0.83  \pm 0.09$ & $1.32  \pm 0.09$ & $0.49  \pm 0.13$\\
$28$ & $1.01  \pm 0.11$ & $1.22  \pm 0.09$ & $0.21  \pm 0.14$\\
$29$ & $1.10  \pm 0.12$ & $1.17  \pm 0.08$ & $0.08  \pm 0.14$\\
$30$ & $1.18  \pm 0.13$ & $1.12  \pm 0.08$ & $0.06  \pm 0.14$\\
$31$ & $1.27  \pm 0.14$ & $1.07  \pm 0.08$ & $0.20  \pm 0.15$\\
$32$ & $1.35  \pm 0.15$ & $1.01  \pm 0.07$ & $0.33  \pm 0.16$\\
$34$ & $1.52  \pm 0.16$ & $0.91  \pm 0.07$ & $0.61  \pm 0.17$\\
\hline
\end{tabular}
\end{center}
\end{table}

\begin{table}
\caption{Critical exponent $\gamma$ for 'gas' and 'liquid' phases for
different choices of trial value of critical point $m_c$ for
energy 4.0 GeV/nucleon used in gamma matching procedure. Non-normalized
moments.}
\label{Tab_gamma4Gev}
\begin{center}
\begin{tabular}{|c|c|c|c|c|}
\hline
trial $m_c$ & $\gamma_{liquid}$ & $\gamma_{gas}$& $|\gamma_{gas} -
\gamma_{liquid}|$ \\
\hline
$26$ & $0.77  \pm 0.07$ & $1.53  \pm 0.08$ &$ 0.75 \pm 0.11$\\
$28$ & $0.96  \pm 0.08$ & $1.42  \pm 0.07$ &$ 0.46 \pm 0.11$\\
$29$ & $1.05  \pm 0.09$ & $1.36  \pm 0.07$ &$ 0.31 \pm 0.11$\\
$30$ & $1.13  \pm 0.09$ & $1.30  \pm 0.07$ &$ 0.17 \pm 0.11$\\
$31$ & $1.22  \pm 0.09$ & $1.24  \pm 0.07$ &$ 0.02 \pm 0.11$\\
$32$ & $1.30  \pm 0.10$ & $1.18  \pm 0.06$ &$ 0.13 \pm 0.12$\\
$34$ & $1.47  \pm 0.11$ & $1.06  \pm 0.06$ &$ 0.41 \pm 0.14$\\
\hline
\end{tabular}
\end{center}
\end{table}

Results of the calculations for energy 10.6 GeV/nucleon are
presented in Tables \ref{Tab_gamma10Gev-norm} and
\ref{Tab_gamma10Gev}. Table \ref{Tab_gamma10Gev-norm} shows
results obtained based on normalized charge distribution moments
whereas Table \ref{Tab_gamma10Gev} present results from
calculation based on non normalized moments. It is easily seen
that both procedures give exponents that agree with each other
within statistical errors. Final value of exponent $\gamma$ is
calculated as a mean value of $\gamma_{gas}$ and $\gamma_{liquid}$
for normalized moments. The critical point is determined at $ m_c
= 31 \pm 2$ and the critical exponent $\gamma = 1.17 \pm 0.09$.

\begin{figure}
\includegraphics[scale=0.48,bb= 13 280  850 810]{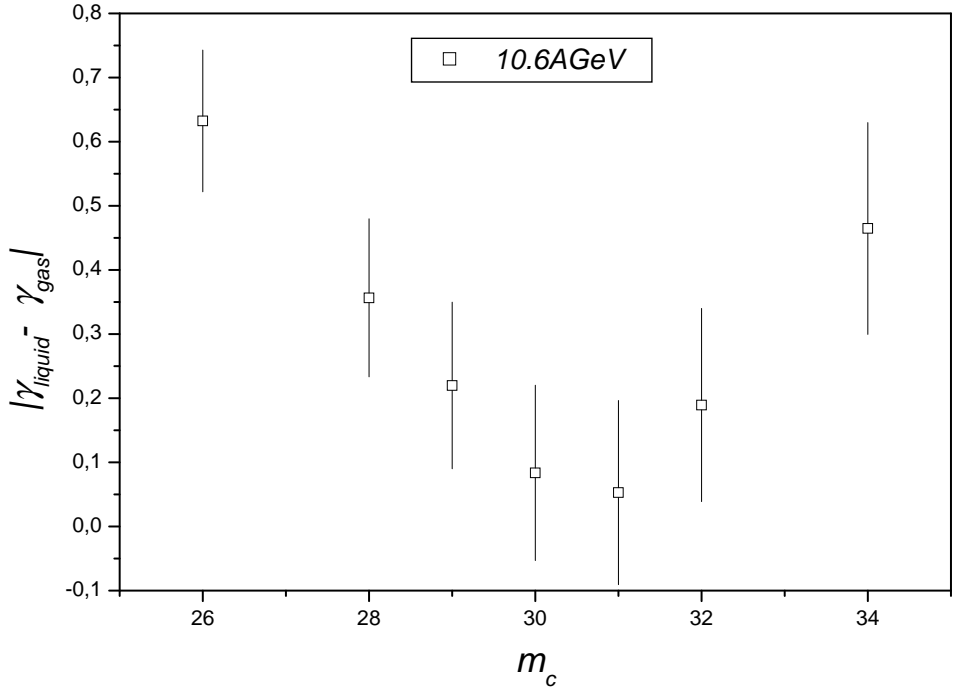}
\caption{Difference $|\gamma_{liquid} - \gamma_{gas}|$ as a
function of trial multiplicity $m_c$ for energy 10.6 GeV/nucleon.}
\label{Fig_6}
\end{figure}

\begin{figure}
\includegraphics[scale=0.48,bb= 13 280  850 810]{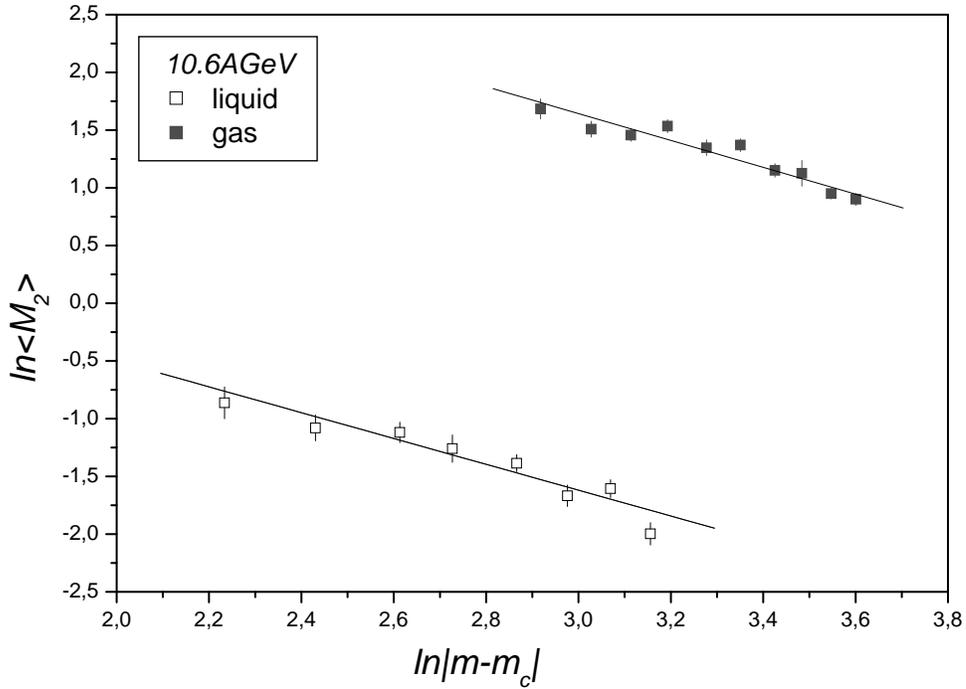}
\caption{Mean value of second charge moment $\langle M_{2}\rangle$
for a 'liquid' and 'gas' as a function of multiplicity $m$ for
energy 10.6 GeV/nucleon. The plotted lines show fits to experimental
data used to determine critical exponent $\gamma$.}
\label{Fig_7}
\end{figure}

The analogous results of the 'gama matching' procedure for energy 4.0
GeV/nucleon are given in Tables \ref{Tab_gamma4Gev-norm} and
\ref{Tab_gamma4Gev}. The determined critical point is $ m_c =
30\pm 2$ and the critical exponent $\gamma = 1.15 \pm 0.09$.
Figure \ref{Fig_6} shows values of $|\gamma_{liquid}
-\gamma_{gas}|$ plotted as a function of trial values of critical multiplicity 
at energy 10.6 Gev/nucleon. Figure \ref{Fig_7} presents results
of fitting procedure for $\gamma_{liquid}$ and $\gamma_{gas}$ for
critical multiplicity value $m_c=30$ for the same energy.

\section{Critical exponents $\beta$ and $\tau$}

\begin{figure}
\includegraphics[scale=0.48,bb= 13 280  850 810]{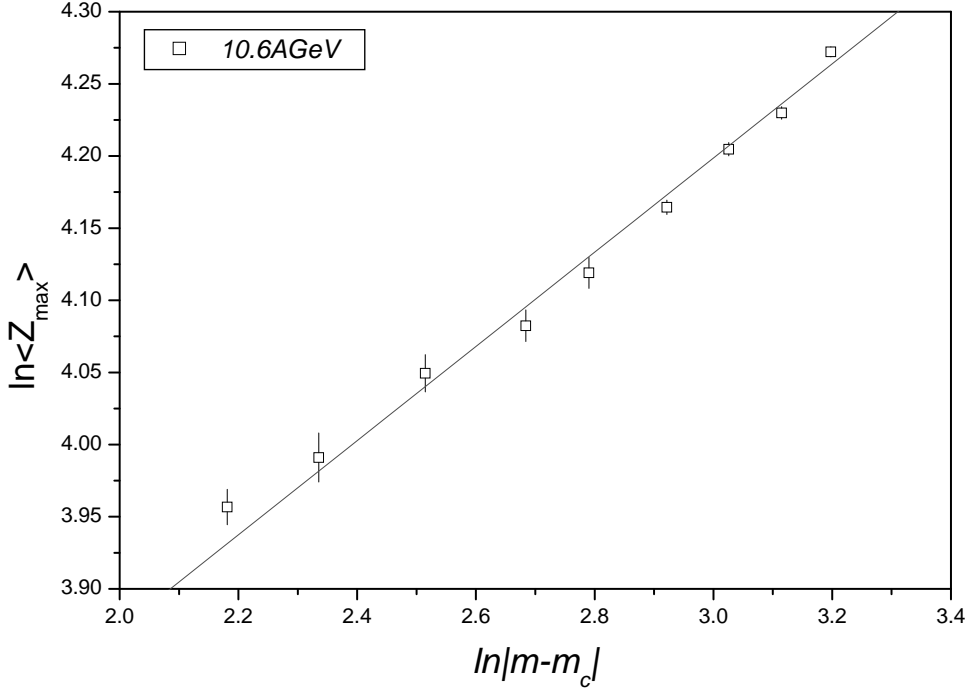}
\caption{Mean Value of  $ln\langle Z_{max}\rangle$ as a function
of distance from critical point $\epsilon$ for 'liquid' phase 10.6
GeV/nucleon.} \label{Fig_11}
\end{figure}

After determination of the critical multiplicity $m_c$ and
ranges of $\epsilon$ in which to fit the critical exponents, determination of 
the
exponents $\beta$ i $\tau$ is
straightforward. Based on Equation \ref{eq:5}, the mean value of
$\ln Z_{max}$ was plotted as a
function of
$\ln |m-m_c|$ (Figure \ref{Fig_11}). The slope of the linear
fit to the plotted relation gives the critical
exponent $\beta$. A linear fit was made within the fitting
boundaries determined during the 'gamma matching procedure'. For
energy 10.6 GeV/nucleon the determined value of $\beta = 0.33 \pm
0.01$ with $\chi^2/ndf = 1.66$, and for 4.0 GeV/nucleon $\beta
= 0.34 \pm 0.01$ with $\chi^2/ndf = 0.99$.

\begin{figure}
\includegraphics[scale=0.48,bb= 13 280  850 810]{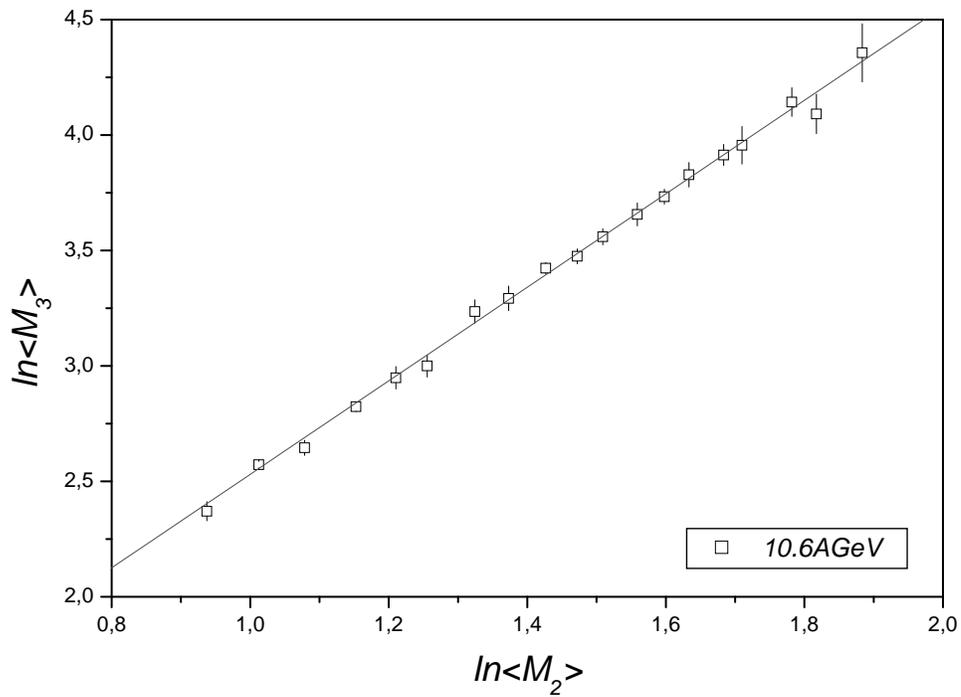}
\caption{Relation between charge moments $\ln \langle
M_{3}\rangle$ and $\ln \langle M_{3}\rangle$ for gas phase 10.6
GeV/nucleon.} \label{Fig_12}
\end{figure}

\begin{figure}
\includegraphics[scale=0.48,bb= 13 280  850 810]{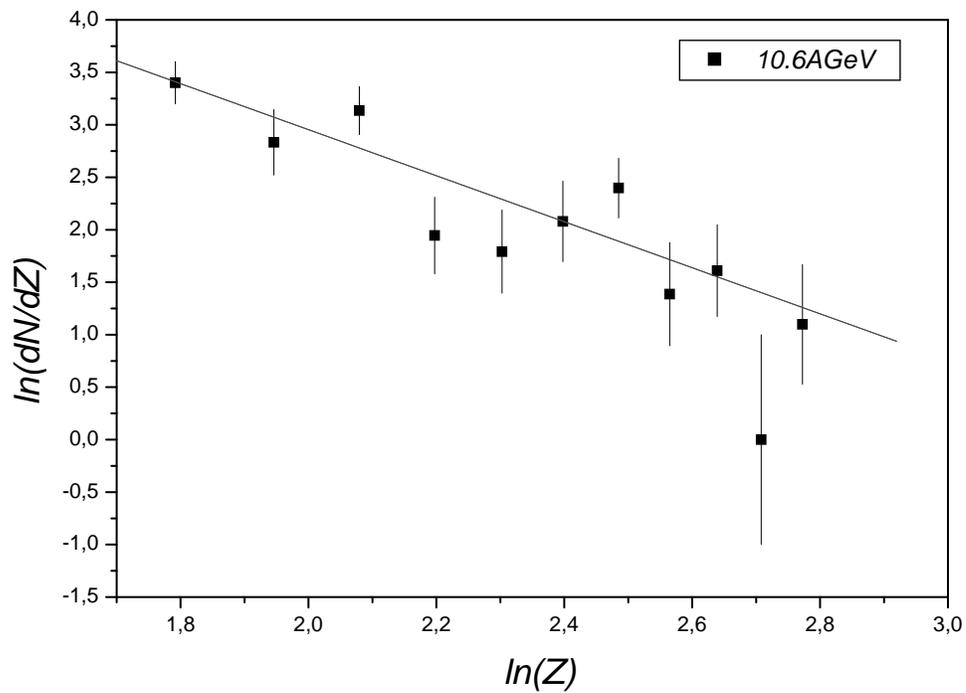}
\caption{Normalized charge distribution for event with
multiplicity $m$ within a range $25 < m < 35$.} \label{Fig_13}
\end{figure}

The critical exponent $\tau$ was determined from equation
\cite{Gilkes94,Elliott011}:
\begin{equation}
\label{eq:3_29}
    \frac{\Delta \ln \langle M_3 \rangle}{ \Delta \ln \langle M_2 \rangle} =
\frac{\tau -4}{\tau
    -3}.
\end{equation}
Figure \ref{Fig_12} presents $\ln \langle M_3 \rangle$ versus $\ln \langle M_2 
\rangle$
 for
the 'gas' phase  
at energy 10.6 GeV/nucleon together with the linear fit.  The slope
of this linear fit is used to determine the $\tau$ exponent, which is  $\tau = 
2.11 \pm 0.05$
for  10.6 GeV/nucleon and  $\tau = 2.12 \pm 0.04$ for energy
4.0 GeV/nucleon. To verify the consistency, the exponent $\tau$
also was determined from Equation \ref{eq:4}. This Equation is supposed to be
valid at the critical point, but in order to have
sufficient statistics, data with $ 25 < m < 35$ were used.
Normalized charge multiplicity distribution was calculated and
plotted in Figure \ref{Fig_13}. The exponent $\tau$ is given by the
slope of the linear fit to the data points for charges from $Z=6$ to
$Z=16$. For charges smaller than 6, the assumptions of the Fisher model
are not valid \cite{Panagiotou85} and for charges larger than 16
experimental data statistics is too small. The resulting value $\tau =
2.19 \pm 0.33$, with reduced $\chi^2 = 1.2$ is in agreement with
previously determined value.

\section{Discussion}

\begin{figure}
\includegraphics[scale=0.48,bb= 13 280  850 810]{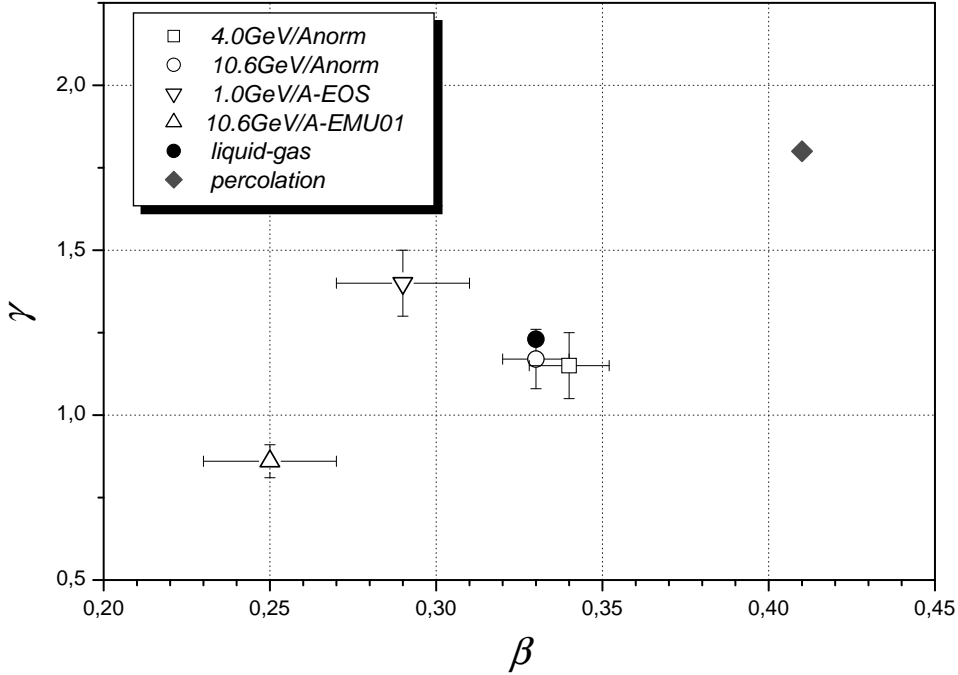}
\caption{Comparison of determined and predicted values of critical
exponent $\gamma$ and $\beta$.} \label{Fig_14}
\end{figure}

Table \ref{Table_exp} summarizes the determined values of critical
exponents $\gamma$, $\beta$ and  $\tau$ at energies 4.0 and  10.6
GeV/nucleon both for normalized and non-normalized charge moments. For 
comparison, exponents obtained by the EOS, EMU01 and KLMM 
experiments are also shown, as well as the values predicted by the liquid-gas 
and percolation models. The data on the $\beta$ and $\gamma$ exponents from 
Table \ref{Table_exp} are plotted in Figure \ref{Fig_14}. 
The values of $\beta$ and $\gamma$ exponents determined in this work are very 
close to those expected for the 'liquid-gas' phase transition, both for 4.0 
and 10.6 GeV/nucleon data.
We note that the critical exponents discussed here were determined from the 
selected 'liquid' and 'gas' groups of events. Earlier analyses, without such a 
selection, gave values of the exponents which do not agree with any of the 
models.

Values of the
critical exponent $\tau$ predicted by the percolation and liquid-gas
models are not too different. As experimental errors are 
large compared to the difference between the predicted values of
$\tau$,  this exponent cannot be used to
discriminate models of  multifragmentation.

\begin{table}
\caption{Critical exponents $\gamma$, $\beta$, $\tau$ for energy
4.0 GeV/nucleon and 10.6GeV/nucleon} \label{Table_exp}
\begin{center}
\begin{tabular}{|c|c|c|c|}
\hline $Data $ & $\gamma$ & $\beta$ & $\tau$ \\
\hline
Au-Em  4.0A GeV norm   & $1.15 \pm 0.09 $&$0.34 \pm 0.01$ &$2.12 \pm 0.04$ \\
Au-Em 10.6A GeV norm   & $1.17 \pm 0.09$&$0.33 \pm 0.01$ &$2.11 \pm 0.05$ \\
Au-Em  4.0A GeV        & $1.23 \pm 0.09$&$0.36 \pm 0.02$ &$2.15 \pm 0.04$ \\
Au-Em 10.6A GeV        & $1.11 \pm 0.09$&$0.33 \pm 0.02$ &$2.16 \pm 0.05$ \\
\hline
Au-C   1.0A GeV (EOS \cite{Elliott011})    & $1.40 \pm 0.10$&$0.29 \pm 0.02$ 
&$2.14 \pm 0.06$ \\
Au-Em 10.6A GeV (EMU01 \cite{Adamovitch99})& $0.86 \pm 0.05$&$0.25 \pm 0.02$ 
&$2.23 \pm 0.05$ \\
Au-Em 10.6A GeV (KLMM \cite{Cherry95})     & $              $&$0.19 \pm 0.02$ 
&$1.88 \pm 0.06$ \\
\hline
Liquid-gas model & $1.23$&$0.33$ &$2.21$ \\
Percolation model & $1.80$&$0.41$&$2.18$ \\
\hline
\end{tabular}
\end{center}
\end{table}

Presented in this work a suggestion that the Fisher's liquid drop model
properly describes multifragmentation process, is in agreement with
recently published results of the ISiS collaboration
\cite{Elliott01_2}, confirming that the Fisher's scaling law is
followed by experimental data. In addition, Mader et al.
\cite{Mader01} have shown a similarity between predictions of
the Fisher model and clusterisation process in the three-dimensional
Ising model. Therefore, multifragmentation can be interpreted in a
similar way as condensation of liquid drops in equilibrium with
the bulk liquid.

One can try to estimate a critical temperature of the system by using
a relation between multiplicity and temperature of the system given
in  \cite{Hauger96,Hauger98}. Estimation of the initial
temperature of the system after collision, $T_i$, was based on
Fermi gas model and does not take into account the expansion of the
system. The final temperature of the system at breakup, $T_f$, was
estimated based on a relation between multiplicity $m$ and
temperature $T$ for isotopic-yield-ratio thermometer given in
\cite{Hauger96}. Our values of critical
 multiplicity ($m_c=30$ and $m_c=31$ at 4.0 and 10.6 GeV/nucleon, 
respectively) correspond to $T_i = 9.3
\pm 0.7$ MeV and $T_f = 5.4 \pm 0.2$ MeV. The error of $T_i$
reflects  uncertainty of the input parameters in the Fermi gas
model. These $T_i$ and $T_f$ values bracket the critical temperature of the
system.  This estimated range for the critical temperature agrees with
predictions of the Fisher model if we take into account the
scaling of  critical temperature with the finite system size
\cite{Elliott011}. The estimation of the critical temperature
relies on the assumption that relation between multiplicity $m$
and temperature of the system established for energy 1.0 A GeV applies
also for higher energies 4.0 and 10.6 GeV/nucleon.

The above results, together with results of ISiS collaboration
\cite{Elliott01_2} suggest that not all multifragmentation events
undergo a second order phase transition. The data excluded
from our analysis presented above may be interpreted as events in which
multifragmentation occurs far away from the critical point. These
events could undergo a first order phase transition so that 'gas' and
'liquid' coexist inside the nucleus as suggested by experimental
analysis of GSI data \cite{Pochodzalla95}.

As noted earlier, the excited nucleus after a collision evolves into the 
neighborhood of
the critical point on the density--temperature plane.
If multifragmentation occurs close to the
critical point, critical exponents characterizing second order
phase transition occurring in the neighborhood of the critical
point should be observed. If for some events multifragmenation
occurs far away from the critical point and they are included in the
analysis, the observed values of critical exponents could be altered so
that the whole picture of the physical process is obscured. In
such situations proper selection of events should help  obtain
true values of critical exponents.

Events excluded from our analysis show very strong fluctuations of
the second moment of the charge distribution significantly far
away from the critical point, especially in the liquid phase. Large
fluctuations could result from coexistence of two different
phases with different properties inside a single nucleus. Below
critical multiplicity $m_c$ (proportional to temperature of the
system) two separate phases with significantly different
properties could coexist in nucleus as is suggested by mean field
\cite{De97,De99,Sil01} or canonical model calculations
\cite{Lee01}. The coexistence of the two phases may take place in wide
ranges of temperature and pressure. This could explain large
fluctuations of second moment of charge distribution far away from
the critical point.  It is also important to stress that moment $M_2$
is proportional to the isothermal compressibility of the system.
Therefore, large fluctuations of $M_2$
correspond to fluctuations of the compressibility $\kappa_T$ and
to fluctuation of the density of the system. Large fluctuations of
$M_2$ observed in 'liquid phase' (Figure \ref{Fig_3}) may be
interpreted as resulting from large density difference between the two
coexisting phases.\\

In summary, an attempt to extract critical exponents $\gamma$,
$\beta$ and $\tau$ was performed, using data coming from
interactions of gold nuclei with nuclear emulsion at
energies 4.0 A GeV and 10.6 A GeV.
To extract the exponents, two subsets of data with characteristics
similar to that of 'gas' and 'liquid' phases were selected, based on analysis
of Campi's 2nd charge moments. The extracted values of the critical
exponents for the selected data sets are in agreement with
predictions of liquid-gas model of phase transition. The same
analysis performed without the selection of 'gas' and 'liquid' samples
favors neither percolation nor liquid-gas model of phase
transition.  A suggestion is made that data  excluded from the
above mentioned samples represents events where phase transition, 
if any,  occurs far from the critical point.

\end{document}